\documentclass[10pt]{article}
\usepackage{epsfig}
\usepackage{amssymb}
\usepackage{amsmath}

\newcommand{\ud}{\mathrm{d}}

\begin{document}

\title{\bf Sigma Model Q-balls and Q-Stars}

   \author{
  \large Y. Verbin }
 \date{ }
   \maketitle
       \centerline{\em Department of Natural Sciences, The Open University
   of Israel,}
   \centerline{\em P.O.B. 39328, Tel Aviv 61392, Israel}
  \centerline{ e-mail: \em verbin@openu.ac.il}
   \vskip 1.1cm

\begin{abstract}
 A new kind of Q-balls is found: Q-balls in a non-linear sigma model. Their main properties are presented together 
 with those of  their self-gravitating generalization, sigma model Q-stars. A simple special limit of solutions 
 which are bound by gravity alone (``sigma stars'') is also discussed briefly. 
The analysis is based on calculating the mass, global U(1) charge and binding energy for families of solutions 
parameterized by the central value of the scalar field. 
Two kinds (differing by the potential term) of  the new sigma model Q-balls and Q-stars are analyzed. They are 
found to share some characteristics while differing in other respects like their properties for weak central 
scalar fields which depend strongly on the form of the potential term. 
They are also compared with their ``ordinary'' counterparts and although similar in some respects, 
significant differences are found like  the existence of an upper bound on the central scalar field. 
The sigma model Q-stars also contain non-solitonic solutions whose relation with sigma star solutions is discussed. 

\end{abstract}

\section{Introduction}
\setcounter{equation}{0}

  Q-balls \cite{Coleman1985} occur in a wide variety of (theoretical) physical contexts. They appear naturally in
the minimal supersymmetric Standard Model \cite{Kusenko1997b,EnqvistMacD1998}
as condensates of squarks or sleptons. The larger ones ($Q \gtrsim 10^{15}$)
can have a crucial cosmological significance as dark matter candidates \cite{KusenkoShaposh1998}
if they are stable or long living, or (since they carry baryon or lepton number)
as a possible explanation for the baryon asymmetry in the universe
\cite{EnqvistMacD1998} and the baryon to dark matter ratio \cite{EnqvistMacD1999}.
Small Q-balls \cite{Kusenko1997a} can be produced even more easily in high temperatures
and may also be found as dark matter. They may also be produced
in colliders for direct inspection of their interesting properties
 \cite{Kusenko1997b,DvaliEtAl1997}. See also e.g. Enqvist\&Mazumdar\cite{EnqvistMazumdar2003} 
 and Dine\&Kusenko \cite{DineKusenko2004} for further reviews.
 A large number of discussions of various other aspects of 
 Q-balls exists already with different approaches: analytic
 \cite{Kusenko1997a,CorreiaSchmidt2001,IoannidouEtAl2005a,IoannidouEtAl2005b},
mixed -  analytic and numerical 
 \cite{MultamakiVilja2000,IoannidouEtAl2003,VolkovWohnert,KleihausEtAl2005,Verbin2007},
 numerical simulations  \cite{AxenidesEtAl2000,BattyeSutcliff2000} for addressing more complicated 
 issues like  scattering (not yet in 3 spatial dimensions) and so on.

All the above-mentioned Q-ball studies are based on the ``original'' flat space Q-balls.
However, it is evident that for a large enough mass scale, gravitational effects become
important and one needs to study Q-stars \cite{Jetzer1992,LeePang1992, Liddle1992}. 
The existence of Q-stars was demonstrated by Friedberg {\it et al} \cite{FriedbergLeePang1986b}
and by Lynn \cite{Lynn89}. Further studies revealed more features like the fact that gravity 
limits the size of Q-balls  \cite{MultamakiVilja2002} or the properties of spinning 
 Q-stars \cite{KleihausEtAl2005}, or made generalizations like  Q-stars with non-minimal coupling
 to gravity \cite{Prikas2002}.

Furthermore, unifying theories (typically in higher dimensions) lead frequently to 
non-linear sigma models, so Q-balls and Q-stars should be studied in these models as well. 
Although topological solitons in non-linear sigma models  \cite{Rajaraman} have been studied for 
decades, it seems that non-topological solitons of the same models have received very
little attention. This work will be devoted to a special type of those:
Q-balls and their self-gravitating counterparts, i.e. sigma model Q-stars. 

\section{General Considerations}
\setcounter{equation}{0}

Sigma model Q-balls and Q-stars are spherically symmetric solutions of the field equations derived from 
the action
\begin{eqnarray}
S=\int \ud^4x\sqrt{|g|}\left( \frac{1}{2}{\cal E }(|\Phi|)(\nabla_\mu \Phi)^*(\nabla ^\mu
\Phi)-U(|\Phi|)+\frac{1}{16\pi {\cal G}}R\right)
\label{action1}
\end{eqnarray}
where ${\cal E }(|\Phi|)$ is a non-negative dimensionless function, which may be 
interpreted as a Weyl factor of a conformally-flat (two-dimensional) target space metric. 
A particularly simple system which will be studied here is the  O(3) sigma model \cite{Rajaraman}
which corresponds to ${\cal E }(|\Phi|) = 1/(1+|\Phi|^2/m^2)^2$. This is the conformal factor of
a target space of $\mbox{\textrm{S}}^2$ with a diameter $m$.

The function $U(|\Phi|)$ is a non-negative potential which will be chosen
such as Q-ball solutions will exist as will be explained below. 
More notations and conventions: $\nabla_\mu$ is the covariant derivative, the signature
is $(+,-,-,-)$ and $R^\kappa_{ \lambda\mu\nu}
=\partial_\nu\Gamma^\kappa_{\lambda\mu} -
\partial_\mu\Gamma^\kappa_{\lambda\nu}+ \cdot\cdot\cdot$.
${\cal G}$ is Newton's constant and $G_{\mu\nu}$ will denote the Einstein tensor.

The field equations derived from the action (\ref{action1}) are:
\begin{equation}
{\cal E }(|\Phi|)\nabla_\mu \nabla^\mu \Phi +  
\frac{\Phi^*}{2|\Phi|}\frac{d\cal E}{d|\Phi|}\nabla_\mu \Phi\nabla ^\mu \Phi+
\frac{\Phi}{|\Phi|} \frac{dU}{d|\Phi|} = 0
\label{ScalarFEeq}
\end{equation}
and
\begin{eqnarray}
-\frac{1}{8\pi {\cal G}} G_{\mu\nu}= T_{\mu\nu} =
 \frac{1}{2} {\cal E }(|\Phi|) \left[ (\partial_\mu
\Phi)^*(\partial_\nu\Phi)+(\partial_\nu \Phi)^*(\partial_\mu\Phi)
-(\nabla_\lambda \Phi)^*(\nabla^\lambda \Phi) g_{\mu\nu}\right]
+U(|\Phi|)g_{\mu\nu} \label{Einstv1}
\end{eqnarray}
which is equivalent to
\begin{eqnarray}
\frac{1}{8\pi {\cal G}}R_{\mu\nu}+\frac{1}{2} {\cal E }(|\Phi|)\left[
(\partial_\mu \Phi)^*(\partial_\nu\Phi)+(\partial_\nu
\Phi)^*(\partial_\mu\Phi) \right]-U(|\Phi|)g_{\mu\nu} = 0
\label{Einstv2}
\end{eqnarray}

It is well-known that in flat space and a proper choice of potential which ``contains
any attractive interaction however weak'' \cite{LeePang1992}, the ``linear'' system 
$({\cal E }(|\Phi|)=1)$ has non-topological solitons stabilized by the global U(1) charge $Q$.  
The U(1) current density in the general case is given by the slightly modified expression
\begin{equation}
j_{\nu} = -\frac {i}{2}{\cal E }(|\Phi|)(\Phi^*\partial_{\nu}\Phi - \Phi
\partial_{\nu}\Phi^*) \label{current}
\end{equation}
The simplest way to obtain non-vanishing U(1) charge is to allow a uniform rotation in target space
(``field space''),
\begin{equation}
\Phi=F(x^{k})e^{i\omega t} \label{QballAns}
\end{equation}
and indeed it can be proven \cite{LeePang1992,Kusenko1997a} that this must be the form of the
field which minimizes the energy within the sector of a given $Q$ in the ``linear'' theory. The 
generalization to the ``non-linear'' case is straightforward. The condition of finite charge
leads to the boundary condition  $F(x^{k})\rightarrow 0$ at infinity.

If we assume further spherical symmetry, $F(x^{k})$ will
depend only on the radial coordinate $r$ and the line element will take the usual form
\begin{equation}
ds^2=A^2(r)dt^2- B^2(r)dr^2-r^2(d\theta^2+\sin^2\theta d\varphi^2)
\label{staticSphMetric}
\end{equation}
Note that the $e^{i\omega t}$ factor indeed modifies the energy-momentum tensor but
keeps it static.

Einstein equations (\ref{Einstv2}) for Q-stars become
\begin{eqnarray}
\left( \frac{r^2 A'}{B}\right)' =8\pi {\cal G} ABr^2
\left(\frac{\omega^2 {\cal E }(F) F^2}{A^2} -U(F)\right)
\label{EinsEqQballsR00} \\
\frac {A''}{A} - \frac{A'B'}{AB} - \frac {2B'}{Br}=-8\pi {\cal G}
B^2 \left(\frac{{\cal E }(F) F'^2 }{B^2} +U(F)\right)
\label{EinsEqQballsR11} \\
\frac{1}{r^2}-\frac{1}{B^2 r^2} - \frac {1}{B^2 r} \left(\frac
{A'}{A}-\frac {B'}{B}\right)=8\pi {\cal G} U(F)
\label{EinsEqQballsRang}
\end{eqnarray}
and they should be supplemented by the scalar field equation:
\begin{eqnarray}
{\cal E }(F)\left(\frac{F''}{B^2}+ \frac {(Ar^2/B)'F'}{ABr^2} +
\frac{\omega^2 F}{A^2}\right) + 
\frac{1}{2} \frac{d{\cal E }}{dF}\left( \frac{\omega^2 F}{A^2}+\frac{F'}{B^2} \right)
- \frac{dU}{dF}=0 \label{ScalarEqQball}
\end{eqnarray}

The charge and mass are given by
\begin{eqnarray}
Q=4\pi \omega \int_{0}^\infty dr r^2 (B/A) {\cal E }(F)F^2
\label{Charge} \\
M=4\pi  \int_{0}^\infty dr r^2
\left({\cal E }(F)\left(\frac{\omega^2 F^2}{2A^2} +\frac{F'^2}{2B^2}\right)+U(F)\right)
\label{Mass}
\end{eqnarray}
We note that the sigma Q-star can be viewed as a bound state of $|Q|$ elementary
bosons (that is, it is stable against decay into free bosons) if $M/m <|Q|$. Without loss 
of generality we will assume $\omega>0$ so we will have $Q>0$ as well.

Actually, we have to solve a system of three differential
equations: eq. (\ref{ScalarEqQball}) and only two of the three eqs
(\ref{EinsEqQballsR00})-(\ref{EinsEqQballsRang}). By taking
combinations of those three we get two first order equations
(which may be obtained directly from the $G_{\mu\nu}$ equations). 

It is more comfortable and efficient to introduce a dimensionless mass
function ${\cal M}(r)$ and use it instead of the metric function $B(r)$ following the definition
\begin{equation}
1-\frac{1}{B^2} = \frac{2{\cal G}M(r)}{r}=\frac{2{\cal M}(r)}{mr}
\end{equation}
Note that $M(r)$ is the accumulated mass up to radial coordinate $r$ and the total mass
of the Q-star is the limit $M(\infty)$ which we simply abbreviate by $M$ where there is
no danger for ambiguity. Since the solutions are localized, $M(r)$ becomes
essentially constant quite fast (this is one reason for using it instead of $B(r)$), and
using the numerical radial end point instead of infinity is accurate enough. It is also simpler 
and more natural to revert to an angular field $\Theta$ defined by
\begin{equation}
|\Phi| = m \tan (\Theta/2)
\label{ThetaDef}
\end{equation}

It is straightforward to rewrite the field equations in terms of ${\cal M}(r)$ and $\Theta (r)$ and actually
to cast them in a dimensionless form which is ready for numerical solution. We get:

\begin{eqnarray}
{\cal M}' =\gamma x^2\left[\frac{\overline{\omega}^2 \sin^2 (\Theta)}{8A^2} +
\left(1-\frac{2{\cal M}}{x}\right)\frac{\Theta'^2}{8}+u(\Theta)\right]
\label{SigmaEinsEqQballsG00Dmlss} \\
\left(1-\frac{2{\cal M}}{x}\right)\frac {xA'}{A}=
\gamma x^2\left[\frac{\overline{\omega}^2 \sin^2 (\Theta)}{8A^2} +
\left(1-\frac{2{\cal M}}{x}\right)\frac{\Theta'^2}{8}-u(\Theta)\right]+\frac{{\cal M}}{x}
\label{SigmaEinsEqQballsG11Dmlss} \\
\left(1-\frac{2{\cal M}}{x}\right)\Theta''+
2\left(1-\frac{{\cal M}}{x}-\gamma x^2 u(\Theta)\right) \frac {\Theta'}{x} +
\frac{\overline{\omega}^2 \sin (2\Theta)}{2A^2}- 4\frac{du}{d\Theta}=0
\label{SigmaScalarEqQballDmlss}
\end{eqnarray}
where we use a dimensionless potential function $u(\Theta)$ and define $x=mr$, 
$\overline{\omega}=\omega/m$ and $\gamma=4\pi{\cal G}m^2$.

Actually, the flat space limit of this system with a conserved global charge have been 
studied already to some extent under the title ``Q-lumps'' \cite{Leese1991,Ward2003}, which refer to
solutions which carry additional {\it topological} charge. This kind of topological solutions exists 
only in lower dimensionality or at most as stringlike in four dimensions. However, we will be able to find 
flat space (as well as self-gravitating) spherical solutions (in four spacetime dimensions) since we give 
up topological non-triviality. 
Therefore, our solutions may be simply regarded as sigma model Q-balls and Q-stars stabilized by the
global charge alone. Within the context
of the present discussion, the difference between Q-balls and Q-lumps is just a difference in the boundary
conditions: both kinds require $\Theta(\infty)=0$ (for finite charge) but Q-lumps exist for 
 $\Theta(0)=\pi$ while the sigma Q-balls we find need $\Theta(0)<\pi/2$. These boundary conditions
are related to the potential functions and it turns out that Q-balls and Q-stars are easily obtained for
a large family of potentials having a global minimum at $\Theta=0$ and another local one. We will use the
simple form $u(\Theta)=\sin^2 (\Theta)/8- \alpha\sin^p (\Theta)/p$ with $p=3$, $4$ so 
the potentials (whose local minimum is always at $\Theta=\pi/2$) are:
\begin{equation}
u_{23}(\Theta)=\frac{\sin^2 (\Theta)}{8}- \frac{\alpha\sin^3 (\Theta)}{3} \;\;\;\;,\;\;\;\;\;\;
u_{24}(\Theta)=\frac{\sin^2 (\Theta)}{8}- \frac{\alpha\sin^4 (\Theta)}{4} \label{SigmaPotentials}
\end{equation}
We will see that the difference between the corresponding solutions will be analogous to those between the 
2-3-4 and the 2-4-6 potentials of the ``linear'' system \cite{Verbin2007}. 

\begin{figure}[!t]
   \begin{center}
      \includegraphics[width=8cm]{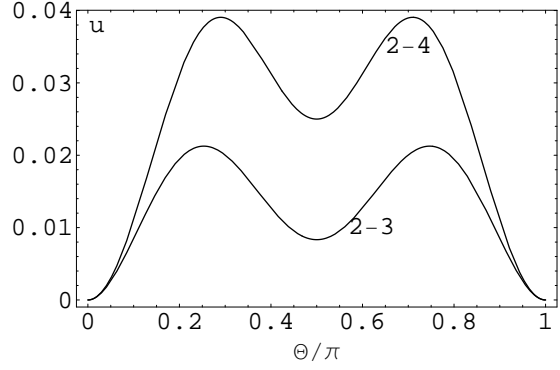} \\
   \caption{Plots of the 2-3 potential for $\alpha=0.35$ and 2-4 potential for $\alpha=0.4$.}
 \label{figureSigmaPot}
     \end{center}
     \end{figure}

The first term in both potentials is just a simple mass term which adds up with the $\omega$ term as 
in the ``linear'' system. The normalization is such that $m$ is still the mass of the elementary free 
scalars. We choose
representative values of $\alpha=0.35$ for the 2-3 potential and $\alpha=0.4$ for the 2-4 one. 
The potentials are shown for these values in figure \ref{figureSigmaPot}. As usual, one may get the 
main properties of the Q-balls from the ``effective potential'' 
$u_{eff} (\Theta)=u(\Theta)-\overline{\omega}^2\sin^2 (\Theta)/8$. The parameter $\overline{\omega}$ 
will take values between $\overline{\omega}_{-}=\sqrt{1-8\alpha/p}$ and 
$\overline{\omega}_{+}=1$ and the corresponding central fields are $\Theta_{\ast}(0)= \pi/2$ and
$\Theta(0)\rightarrow 0$ .

Due to the ``north-south'' symmetry which is left in the potential functions there exists of course 
another family of ``mirror'' solutions with the boundary condition $\Theta(\infty)=\pi$ instead of the usual
$\Theta(\infty)=0$ that we are using. The second boundary condition will satisfy accordingly 
$\Theta(0)>\pi/2$. Obviously, our choice does not lead to any loss of generality of the results that will be 
presented here, but it should be kept in mind that they apply to two different families of solutions. 

\section{Properties of the Solutions} 
\setcounter{equation}{0}

We study solutions with both potentials for the three values 
$\gamma=0$ , $0.02$ , $0.2$. As can be guessed from inspection of the ``effective potential'', the 
``thin wall''  and ``thick wall'' solutions exist for both potentials in a  
way similar to the ``linear'' case.

It is obvious that there are no flat space solutions with $\alpha=0$, that 
is a mass term only. Gravity changes the situation and allows solutions which are the sigma model
analogues of the boson star solutions, so we may call them ``sigma stars''. One difference with respect to 
boson stars is that these sigma stars do not enjoy a scaling symmetry and do not fall on a one-parameter
curve in parameter space. Therefore the $\gamma = 0.2$  family of pure self-gravitating sigma model 
solutions which we present is not universal.

\begin{figure}[!t]
      \begin{center}
       (a) \hspace{65mm} (b)\\
      \includegraphics[width=7.8cm]{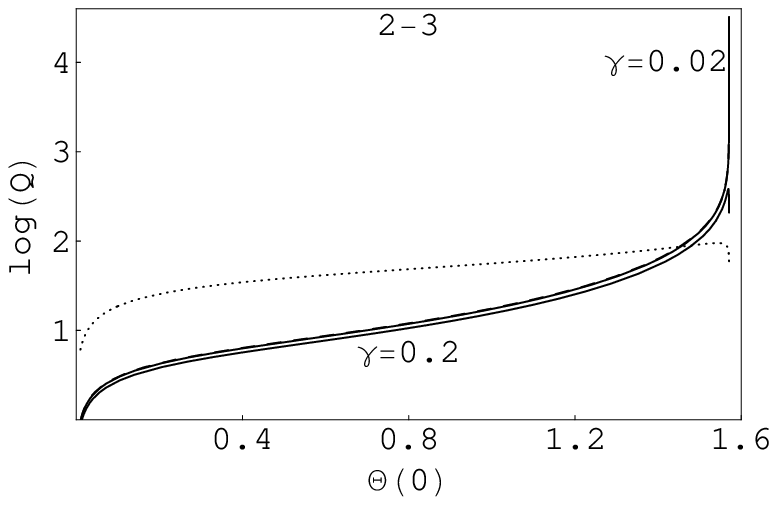}
       \includegraphics[width=7.8cm]{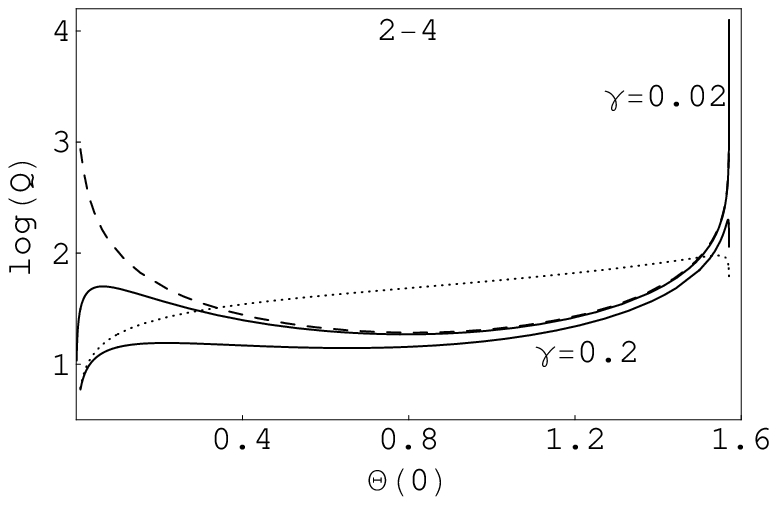}\\
   \caption{Plots of $\log(Q)$  vs.  $\Theta(0)$  for $\gamma=0$ 
   (sigma model Q-balls - dashed),  $\gamma=0.02$ and  $\gamma=0.2$.
   (a) 2-3  sigma model Q-stars; (b) 2-4  sigma model Q-stars. The $\gamma=0$ line cannot be 
   resolved from the $\gamma=0.02$ one in the 2-3 potential.  The dotted lines correspond to 
   sigma stars with $\gamma=0.2$.} 
 \label{figureSigQSLogQ-fx0}
     \end{center}
     \end{figure}

\begin{figure}[!t]
   \begin{center}
     (a) \hspace{65mm} (b)\\
      \includegraphics[width=7.8cm]{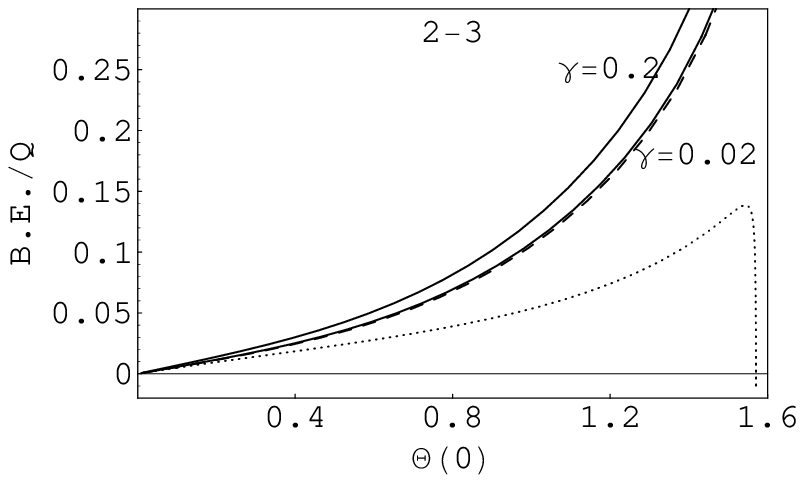}
      \includegraphics[width=7.8cm]{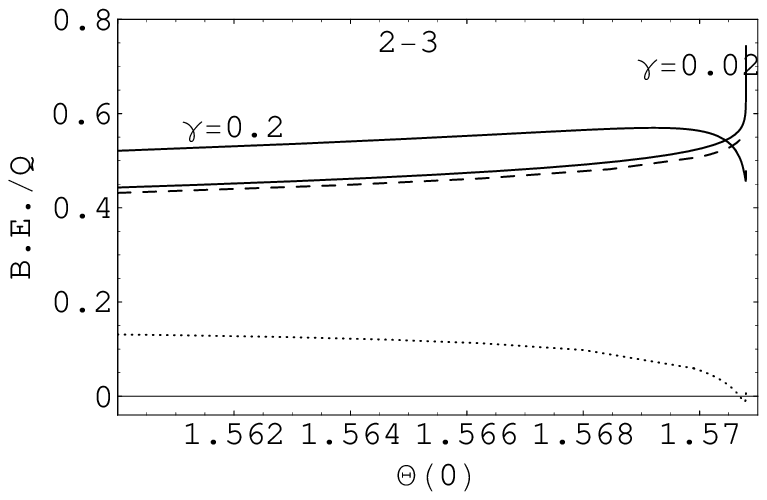}\\
   \caption{Plots of binding energy per particle $(mQ-M)/mQ$ vs. $\Theta(0)$  for 2-3  sigma 
   model Q-stars with $\gamma=0$ (sigma model Q-balls - dashed),  $\gamma=0.02$ and  $\gamma=0.2$.
   (a) $B.E./Q$ up to 0.30; (b) magnification of the large field region with larger $B.E./Q$. 
   The dotted lines correspond to sigma stars with $\gamma=0.2$.} 
 \label{figureSigma23QSBEoverQvsfx0}
     \end{center}
     \end{figure}

\begin{figure}[!t]
   \begin{center}
     (a) \hspace{65mm} (b)\\
      \includegraphics[width=7.8cm]{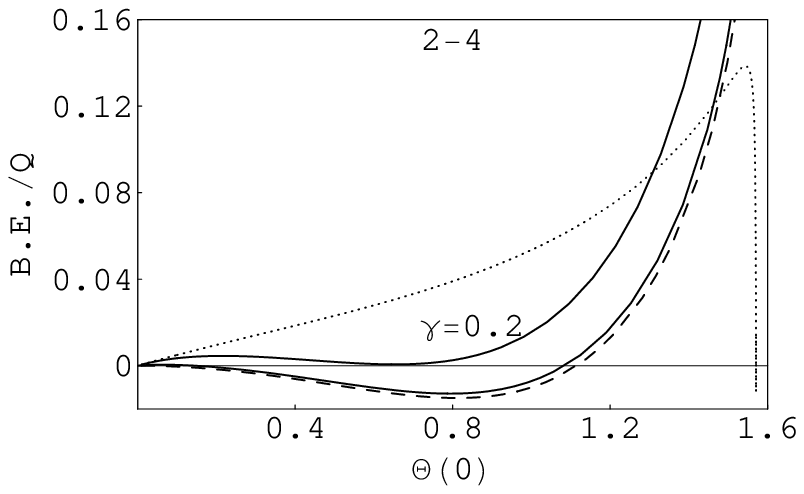}
      \includegraphics[width=7.8cm]{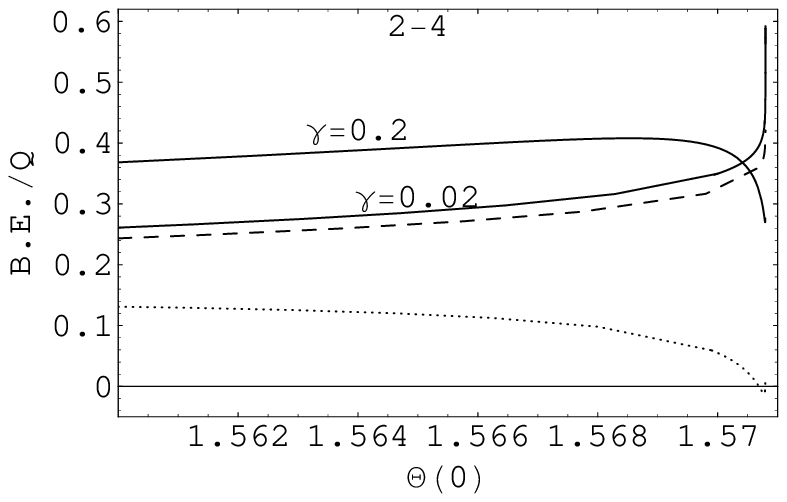}\\
   \caption{Plots of binding energy per particle $(mQ-M)/mQ$ vs. $\Theta(0)$  for 2-4  sigma 
   model Q-stars with $\gamma=0$ (sigma model Q-balls - dashed),  $\gamma=0.02$ and  $\gamma=0.2$.
   (a) $B.E./Q$ up to 0.15; (b) magnification of the large field region with larger $B.E./Q$. 
   The dotted lines correspond to sigma stars with $\gamma=0.2$.} 
 \label{figureSigma24QSBEoverQvsfx0}
     \end{center}
     \end{figure}

\begin{figure}[!t]
   \begin{center}
      \includegraphics[width=7.8cm]{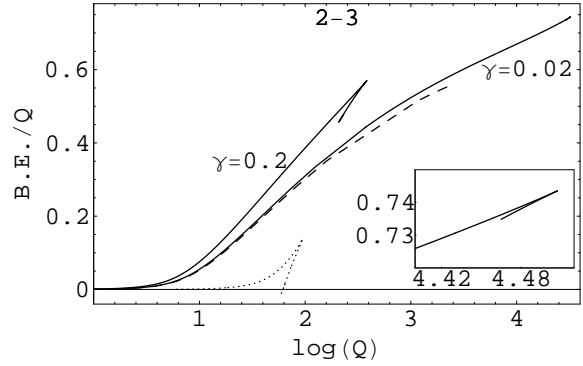}\\
   \caption{Plots of binding energy per particle $(mQ-M)/mQ$ vs. $\log(Q)$  for
   $\gamma=0$ (sigma model Q-balls - dashed), $\gamma=0.02$ and  $\gamma=0.2$ and
   for boson stars with $\gamma=0.2$ (dotted).  The insert is a magnification
   of the upper right corner which contains the $\gamma=0.02$ curve.}
 \label{figureSigma23QSBEoverQvsLogQ}
     \end{center}
     \end{figure}

\begin{figure}[!t]
   \begin{center}
     (a) \hspace{65mm} (b)\\
      \includegraphics[width=7.8cm]{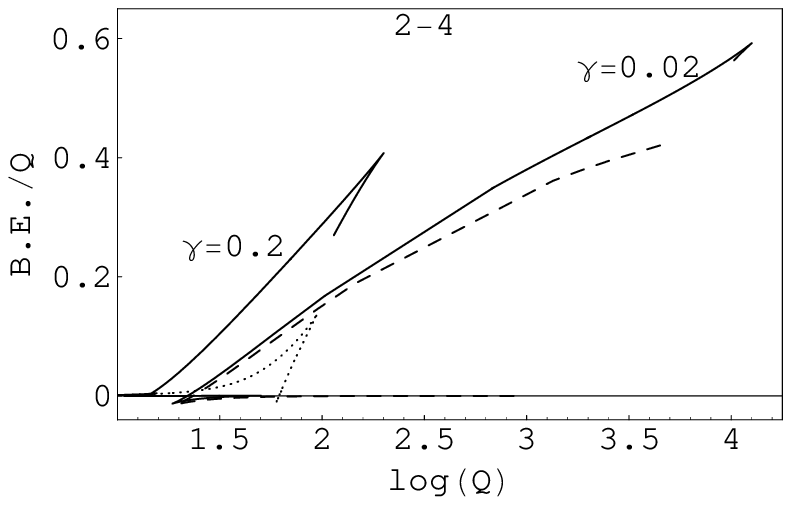}
      \includegraphics[width=7.8cm]{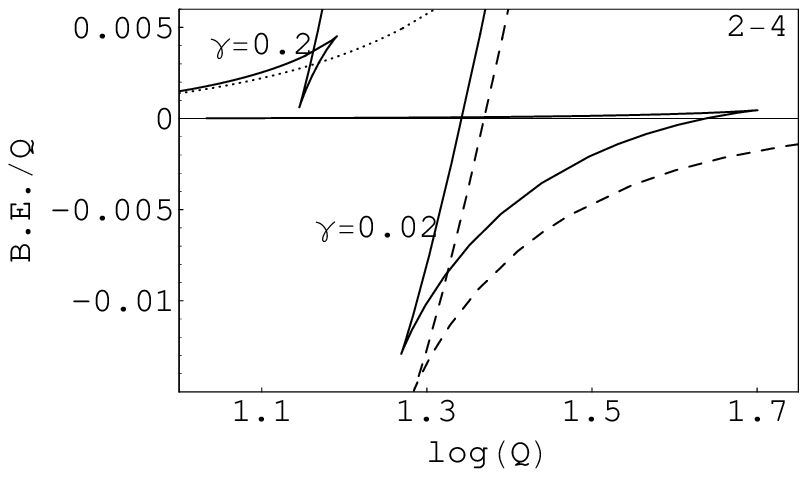}\\
   \caption{Plots of binding energy per particle $(mQ-M)/mQ$ vs. $\log(Q)$  for 2-4  sigma 
   model Q-stars with $\gamma=0$ (sigma model Q-balls - dashed), $\gamma=0.02$ and  $\gamma=0.2$ and
   for boson stars with $\gamma=0.2$ (dotted).  (a) general view; 
   (b) magnification of the small $B.E./Q$ region .}
 \label{figureSigma24QSBEoverQvsLogQ}
     \end{center}
     \end{figure}

The main results are shown in figures \ref{figureSigQSLogQ-fx0}-\ref{figureSigma24QSBEoverQvsLogQ}.
The first, figure \ref{figureSigQSLogQ-fx0} shows the charge as a function of the central 
field $\Theta(0)$ in all cases. It covers the region of maximal charge for $\gamma = 0.02$, although
the maximum is so narrow that its width cannot be seen it the plots. 

It may be expected that curves of the mass as a function of $\Theta(0)$ are superposed in these plots, but 
they are not easily identified in this resolution. However, the plots for the 
binding energy per particle, figures \ref{figureSigma23QSBEoverQvsfx0}-\ref{figureSigma24QSBEoverQvsfx0} 
give enough information about the masses being larger or smaller than $mQ$ for a given $\Theta(0)$. 
More insight into the situation is added by the plots of the binding energy per particle vs. charge, figures 
\ref{figureSigma23QSBEoverQvsLogQ}-\ref{figureSigma24QSBEoverQvsLogQ}.

A new characteristic with respect to the ``linear'' solutions is the limiting value of the central 
scalar field $\Theta(0)$ which is the same for both potentials:  $\Theta_{\ast}(0)= \pi/2$. Moreover, 
 now the same limiting $\Theta(0)$ appears also for the self-gravitating solutions 
and not for Q-balls only. The reason for this is the special form of the potential functions that we 
chose, which yields the minimum of the effective potential to be at $\Theta(0)=\pi/2$ for all values 
of $\omega$. We stress also that $\Theta=\pi/2$ is the equator of $\mbox{\textrm{S}}^2$ and does not
correspond to $|\Phi|\rightarrow \infty$ -- see eq. (\ref{ThetaDef}). Therefore, it is possible that
a different choice of potential function will allow larger values of $\Theta_{\ast}(0)$.

Another new characteristic is the existence of ``mirror'' solutions with different boundary conditions. The
possibility of coexistence of solutions of both type and their interaction deserves further study.

Generally, we find a close parallelism between the main properties of the sigma Q-balls/Q-stars and the analogous
``ordinary'' Q-balls/Q-stars.
The ``damped oscillations'' of $Q$ beyond its maximum which exist in the ``linear'' model \cite{Verbin2007}
exist also here but they are only partially visible in figure \ref{figureSigQSLogQ-fx0} due to 
the $\Theta(0)=\pi/2$ limit. As before the solutions are stable against decay into free bosons only in 
a limited region of $\Theta(0)$ values which corresponds to positive binding energy, or $M/m<Q$. 
Note however that charge degeneracy (two masses or more for the same $Q$) 
exists, so even in this region a higher mass state may decay into a lower state plus additional free bosons 
while conserving $Q$. The other solutions with negative
binding energies may either decay completely into free bosons, or will have a smaller stable boson star
among their decay products.
 
The sigma Q-balls with the 2-3 potential are always stable with ever 
growing charge. Gravity imposes maximal charge values:  for $\gamma=0.2$ the maximum is at $Q=382.97$  
and $\Theta(0)=1.56921$ or $2\Theta(0)/\pi=0.99899$, while for $\gamma=0.02$ the maximum is at $Q=32155.3$. 
The corresponding central scalar field is so close to $\pi/2$ that we give it as  
$1-2\Theta(0)/\pi=1.5543\times 10^{-15}$. 

The sigma star solutions with $\gamma=0.2$ that appear in figure
\ref{figureSigQSLogQ-fx0} have a maximum at $Q=95.07$ and $\Theta(0)=1.54355$ or $2\Theta(0)/\pi=0.98265$.
These values define the region of stable solutions. As figures \ref{figureSigma23QSBEoverQvsfx0} and
\ref{figureSigma23QSBEoverQvsLogQ} show, all the others are either unbound (with 
negative binding energy), or may decay into the stable solutions while conserving particle number.

The 2-3 Q-star curve with $\gamma=0.02$ in figure \ref{figureSigQSLogQ-fx0} is very similar to the $\gamma=0$ 
Q-balls (below the maximum of $Q=32155.3$) and it is impossible to distinguish between them. 
The difference shows up in the binding energy which is shown in figures 
\ref{figureSigma23QSBEoverQvsfx0}a and \ref{figureSigma23QSBEoverQvsLogQ}.

The solutions with the 2-4 potential exhibit a more involved structure which may be described again
according to the central field value. The small $\Theta(0)$ sigma Q-balls (no gravity: $\gamma=0$) 
are large and unstable. The stability region starts at $\Theta(0)=1.11515$ for which $Q=M/m=23.43$
and extends all the way to $\Theta(0)=\pi/2$ with monotonically increasing charge and mass. For
$\gamma=0.02$ the small $\Theta(0)$ behavior changes completely. The small $\Theta(0)$ solutions 
have positive binding energies for $\Theta(0)\leq 0.14476$ (for which $Q=43.37$), passing through a local 
maximum of charge at $\Theta(0)= 0.06400$ and $Q=50.17$. An instability region follows for 
$\Theta(0)\leq 1.08417$. The second range of bound solutions starts at $\Theta(0)= 1.08417$ for which 
$Q=M/m=21.99$ and goes toward $\Theta(0)\rightarrow \pi/2$. There is however a global charge maximum
of $Q=12586.03$ very close to $\Theta(0)= \pi/2$. i.e. such that $1-2\Theta(0)/\pi=1.7686\times 10^{-13}$.
We may therefore conclude that for this 2-4 potential there are stable sigma Q-stars for any charge up to a maximal 
value of $Q=12586.03$ for the parameters we chose. Some charge intervals exhibit charge degeneracy, so the
higher mass solutions will decay.

The solutions for $\gamma=0.2$ have similar behavior with two main quantitative differences: the maximal
charge is now much smaller and has a value of $Q=200.18$ and all solutions have in this case positive 
binding energy due to the stronger gravitational self-attraction.

As far as the flat space limit is concerned, the 2-3 solutions may be divided into two types in accordance
with two regions of the central field interval $0<\Theta(0)<\pi/2$. The sigma Q-stars in most of this interval
 are similar to the flat space ones and the limit $\gamma\rightarrow 0$ (keeping $\Theta(0)$ fixed) gives 
 well-behaved sigma Q-balls. These Q-stars may be classified as solitonic since their existence does not 
 depend upon gravity. 
 The exception is the small region near (and below) $\Theta(0)=\pi/2$ where there are only self-gravitating 
 solutions so they are non-solitonic. The 2-4 solutions may be divided into three types: in the region of 
small $\Theta(0)$ and near $\Theta(0)=\pi/2$ gravity gives rise to (non-solitonic) solutions with no flat 
space limit, while in the medium $\Theta(0)$ values the solutions may be considered solitonic.

\section{Summary and Outlook} \label{Summary and Outlook}
\setcounter{equation}{0}

We found new Q-ball and Q-star solutions in the sigma model system. We presented the main properties of 
the solutions in flat spacetime and for two values of the  gravitational strength $\gamma=0.02$ and $\gamma=0.2$. 
The corresponding characteristics are quite different for the different values of $\gamma$ as is evident from 
the plots of charge vs. central field described above, together with the analysis of the $\Theta(0)$ 
and $Q$ dependence of the binding energy.

We found that the sigma model Q-ball and Q-star properties depend strongly on the form of the 
potential term and that the Q-star solutions split into two main types: one is self-gravitating 
version of the flat space Q-balls, while the other may be identified as non-solitonic solutions which do 
not have a flat space limit.

There is a close parallelism with the ``ordinary'' Q-ball and Q-star solutions \cite{Verbin2007}: 
the correspondence is between the sigma model with the 2-3 potential and the ``linear'' 
system with 2-3-4 potential, and sigma model with the 2-4 potential and the ``linear'' 
system with 2-4-6 potential. However, there are some important differences. The most interesting is the
existence of two ``mirror'' families of solutions differing by their boundary conditions. The
possibility of coexistence of solutions of both types and their interaction was not discussed here 
but deserves further investigation.

Another  issue  which  calls  for  a  further  study  is  that  of  spinning  Q-balls and Q-stars.
Although interesting results \cite{VolkovWohnert,KleihausEtAl2005} already exist for the ``linear'' system,
an analogous study for the new sigma model solutions does not exist. A further more systematic analysis is 
needed in order to clarify questions like the relation between charge, mass and angular momentum of spinning 
Q-stars.

Also needed are a better understanding of the dynamics of instability and decay processes of Q-stars and a study
of their possible gravitational collapse.

\newpage

\begin{thebibliography}{99}


\bibitem{Coleman1985}
  S.~Coleman,
  Nucl.\ Phys.\ B {\bf 262}, 263 (1985)
  [Erratum-ibid.\ B {\bf 269}, 744 (1986)].  \vspace{-0.15cm}
  
 \bibitem{Kusenko1997b}
  A.~Kusenko,
  Phys.\ Lett.\ B {\bf 405}, 108 (1997). \vspace{-0.15cm} 
  
  \bibitem{EnqvistMacD1998}
  K.~Enqvist and J.~McDonald,
  Phys.\ Lett.\ B {\bf 425}, 309 (1998). \vspace{-0.15cm} 
  
\bibitem{KusenkoShaposh1998}
  A.~Kusenko and M.~E.~Shaposhnikov,
  Phys.\ Lett.\ B {\bf 418}, 46 (1998).  \vspace{-0.15cm}
  
  \bibitem{EnqvistMacD1999}
  K.~Enqvist and J.~McDonald,
   Nucl.\ Phys.\ B {\bf 538}, 321 (1999).  \vspace{-0.15cm}

\bibitem{Kusenko1997a}
  A.~Kusenko,
  Phys.\ Lett.\ B {\bf 404}, 285 (1997).  \vspace{-0.15cm}

\bibitem{DvaliEtAl1997}
  G.~R.~Dvali, A.~Kusenko and M.~E.~Shaposhnikov,
  Phys.\ Lett.\ B {\bf 417}, 99 (1998).  \vspace{-0.15cm}

\bibitem{EnqvistMazumdar2003}
  K.~Enqvist and A.~Mazumdar,
  Phys.\ Rept.\  {\bf 380}, 99 (2003).  \vspace{-0.15cm}
  
\bibitem{DineKusenko2004}
  M.~Dine and A.~Kusenko,
  Rev.\ Mod.\ Phys.\  {\bf 76}, 1 (2004). \vspace{-0.15cm}
  
\bibitem{CorreiaSchmidt2001}
  F.~Paccetti Correia and M.~G.~Schmidt,
  Eur.\ Phys.\ J.\ C {\bf 21}, 181 (2001). \vspace{-0.15cm}
  
\bibitem{IoannidouEtAl2005a}
  T.~A.~Ioannidou, A.~Kouiroukidis and N.~D.~Vlachos,
  J.\ Math.\ Phys.\  {\bf 46}, 042306 (2005). \vspace{-0.15cm}

\bibitem{IoannidouEtAl2005b}
  T.~A.~Ioannidou, A.~Kuiroukidis and N.~D.~Vlachos,
   Theor.\ Math.\ Phys.\  {\bf 144}, 1171 (2005)
  [Teor.\ Mat.\ Fiz.\  {\bf 144}, 342 (2005)]. \vspace{-0.15cm}  

\bibitem{MultamakiVilja2000}
  T.~Multamaki and I.~Vilja,
   Nucl.\ Phys.\ B {\bf 574}, 130 (2000). \vspace{-0.15cm}

\bibitem{IoannidouEtAl2003}
  T.~A.~Ioannidou, V.~B.~Kopeliovich and N.~D.~Vlachos,
    Nucl.\ Phys.\ B {\bf 660}, 156 (2003). \vspace{-0.15cm}

\bibitem{VolkovWohnert}
  M.~S.~Volkov and E.~Wohnert,
  Phys.\ Rev.\ D {\bf 66}, 085003 (2002). \vspace{-0.15cm}

\bibitem{KleihausEtAl2005}
  B.~Kleihaus, J.~Kunz and M.~List,
  Phys.\ Rev.\ D {\bf 72}, 064002 (2005). \vspace{-0.15cm}

\bibitem{Verbin2007}
Y. ~Verbin, arXiv: 0708.2673 [gr-qc], to be published in the proceedings of the 11th Marcel Grossmann Meeting,
  Berlin, July 2006. \vspace{-0.15cm} 
  
\bibitem{AxenidesEtAl2000}
  M.~Axenides, S.~Komineas, L.~Perivolaropoulos and M.~Floratos,
  Phys.\ Rev.\ D {\bf 61}, 085006 (2000). \vspace{-0.15cm}

\bibitem{BattyeSutcliff2000}
  R.~Battye and P.~Sutcliffe,
    Nucl.\ Phys.\ B {\bf 590}, 329 (2000). \vspace{-0.15cm}   
  
\bibitem{Jetzer1992}
  P.~Jetzer,
  Phys.\ Rept.\  {\bf 220}, 163 (1992). \vspace{-0.15cm}
  
\bibitem{LeePang1992}
  T.~D.~Lee and Y.~Pang,
  Phys.\ Rept.\  {\bf 221}, 251 (1992). \vspace{-0.15cm}  

\bibitem{Liddle1992} A.~R.~Liddle and
M.~S.~Madsen, Int.\ J.\ Mod.\ Phys.\ D {\bf 1}, 101 (1992). \vspace{-0.15cm}

\bibitem{FriedbergLeePang1986b}
  R.~Friedberg, T.~D.~Lee and Y.~Pang,
  Phys.\ Rev.\ D {\bf 35}, 3640 (1987). \vspace{-0.15cm} 
  
\bibitem{Lynn89}
B.~W.~Lynn, Nucl.\ Phys.\ B {\bf 321}, 465 (1989). \vspace{-0.15cm} 

\bibitem{MultamakiVilja2002}
T.~Multamaki and I.~Vilja, Phys.\ Lett.\ B {\bf 542}, 137 (2002). \vspace{-0.15cm}

\bibitem{Prikas2002}
  A.~Prikas,
  Phys.\ Rev.\ D {\bf 66}, 025023 (2002). \vspace{-0.15cm}

\bibitem{Rajaraman} R. Rajaraman, \textit{solitons and
Instantons}, (North Holland, Amsterdam 1982). \vspace{-0.15cm}  

\bibitem{Leese1991}
R.~A.~Leese, Nucl.\ Phys.\ B {\bf 366}, 283 (1991). \vspace{-0.15cm}

\bibitem{Ward2003}
R.~S.~Ward, J.\ Math.\ Phys.\  {\bf 44}, 3555 (2003). \vspace{-0.15cm}

\end{thebibliography}
\end{document}